\documentclass[twocolumn,showkeys,aps,prd,showpacs]{revtex4-1}
\UseRawInputEncoding
\usepackage{graphicx}
\usepackage{pifont}
\usepackage[CJKbookmarks,dvipdfm,colorlinks,linkcolor=red,citecolor=blue]{hyperref}
\usepackage{graphicx}
\usepackage{amsmath, amssymb}
\usepackage{makecell}
\usepackage{multirow}
\usepackage{CJK}
\usepackage{mathrsfs}
\usepackage{bm}
\usepackage{amsmath}
\usepackage{dcolumn}
\usepackage{epstopdf}
\usepackage{dsfont}
\usepackage{amssymb}
\usepackage{tabularx}
\usepackage{array}
\usepackage{float}
\usepackage{color}
\usepackage{epstopdf}
\usepackage{mathrsfs}
\usepackage{extarrows}

\begin{document}

\title{Large  spontaneous valley polarization and anomalous valley Hall effect in antiferromagnetic monolayer  $\mathrm{Fe_2CF_2}$}

\author{San-Dong Guo$^{1}$}
\email{sandongyuwang@163.com}
\author{Liguo Zhang$^{1}$ and Yiwen Zhang$^{1}$ and  Guangzhao Wang$^{2}$}
\affiliation{$^1$School of Electronic Engineering, Xi'an University of Posts and Telecommunications, Xi'an 710121, China}
\affiliation{$^2$Key Laboratory of Extraordinary Bond Engineering and Advanced Materials Technology of Chongqing, School of Electronic Information Engineering, Yangtze Normal University, Chongqing 408100, China}
\begin{abstract}
Superior to ferromagnetic (FM) materials, antiferromagnetic (AFM) materials do
not have any net magnetic moment and are robust to external
magnetic perturbation with ultra-high dynamic speed. To achieve spontaneous valley polarization and anomalous valley Hall effect (AVHE) in AFM materials is of great significance for potential
applications in spintronics and  valleytronics.  Here, we predict  an A-type AFM  monolayer  $\mathrm{Fe_2CF_2}$ with large spontaneous valley polarization.  Monolayer  $\mathrm{Fe_2CF_2}$ has zero Berry curvature in momentum space but non-zero
layer-locked hidden Berry curvature in real space, which provides the basic conditions for the realization of AVHE.
 Because $\mathrm{Fe_2CF_2}$ possesses the combined symmetry ($PT$ symmetry) of spatial
inversion ($P$) and time reversal ($T$) symmetry, the spin  is  degenerate, which prevents the AVHE.
An out-of-plane external electric field can be used to produce spin splitting due to the introduction of layer-dependent electrostatic potential, and then layer-locked AVHE can be realized in $\mathrm{Fe_2CF_2}$.
Moreover, the spin order of  spin splitting can be reversed, when  the direction of electric field  is reversed.
It is proved that  the AVHE can be achieved in Janus $\mathrm{Fe_2CFCl}$ without external electric field due to intrinsic built-in electric field. Our works provide an AFM monolayer with excellent properties to realize AVHE.

\end{abstract}

\maketitle

\section{Introduction}
 The  valley is  defined as  the energy extrema of conduction and/or valence bands in momentum
space, which  is regarded as a
degree of freedom  in analogy to charge and spin. The valley index could be used for processing information and performing logic operations  with low power consumption and high speed, known as the concept of valleytronics\cite{q1,q2,q3,q4}.
The rise of two-dimensional (2D) materials  provides unprecedented opportunities for
valley-relevant physics, and the significant advancements have been made\cite{q1,q2,q3,q4}.
As the typical valleytronic
materials, the transition-metal dichalcogenide  (TMD) monolayers  possess  a pair of
degenerate but inequivalent -K and K valleys in the reciprocal space, which exhibit opposite Berry curvature and selective absorption of
chiral light\cite{q8-1,q8-2,q8-3,q9-1,q9-2,q9-3}. When the  spin-orbit coupling (SOC) is considered,  the
 -K and K valleys show opposite spin splitting, which can be called spin-valley locking.
Unfortunately, the spontaneous valley polarization disappears in these nonmagnetic TMD monolayers,  hindering further extension of valleytronics.
Some extrinsic methods have been used to achieve valley splitting, such as external magnetic field, proximity effect, light
excitation\cite{v5,v7,v9,v10,v11}.

The ferrovalley semiconductor (FVS) with intrinsic valley polarization  has been proposed\cite{q10}, which spurs the continued advancement of valleytronics\cite{q11,q12,q13,q13-1,q14,q14-1,q14-2,q15,q16,q17,q18}.
The FVS is generally a hexagonal  ferromagnetic (FM) monolayer with broken spatial inversion symmetry ($P$), which can be used to achieve the  anomalous valley Hall effect (AVHE).
Compared to FM materials, the antiferromagnetic (AFM) materials possess  the high storage density, robustness against external magnetic field, and ultrafast writing speed\cite{v12}. Therefore,  realizing  valley polarization and AVHE in AFM materials
is more meaningful for energy-efficient and ultrafast   valleytronic devices.
A large number of AFM materials with spontaneous valley polarization are built on bilayer, which is constructed using the ferrovalley monolayer  via van der Waals (vdW) interaction\cite{avh1,avh2,avh3,avh4,avh5,avh6}. In these AFM  bilayers,  the AVHE can be achieved, and the spin splitting can be induced by sliding ferroelectric polarization.

It may be more interesting to search for monolayer AFM materials to achieve spontaneous valley polarization and AVHE. For hexagonal AFM monolayer, the spontaneous valley polarization can appear, when including SOC.
However, due to the absence of spin splitting  in the band
structures, the AVHE is prevented.  The spin splitting in AFM materials can be produced by making the magnetic atoms with opposite spin polarization locating in the different environment (surrounding atomic arrangement)\cite{gsd}.
For example, the nonuniform
potential caused by stacking
AFM monolayer $\mathrm{MnPSe_3}$ or $\mathrm{Cr_2CH_2}$ on ferroelectric monolayer $\mathrm{Sc_2CO_2}$ and constructing AFM Janus $\mathrm{Mn_2P_2X_3Y_3}$ (X, Y=S, Se Te; X$\neq$Y) monolayers  produces spin splitting in these AFM monolayers, and then the AVHE can be achieved\cite{v13,v14,v15}. The introduction of layer-dependent electrostatic potential
caused by out-of-plane external electric field can also induce spin splitting in A-type AFM hexagonal monolayer $\mathrm{Cr_2CH_2}$ and tetragonal monolayer $\mathrm{Fe_2BrMgP}$, and the spin order of  spin splitting can be reversed, when flipping  the direction of electric field\cite{gsd1,gsd2}.

\begin{figure}
  % Requires \usepackage{graphicx}
  \includegraphics[width=8cm]{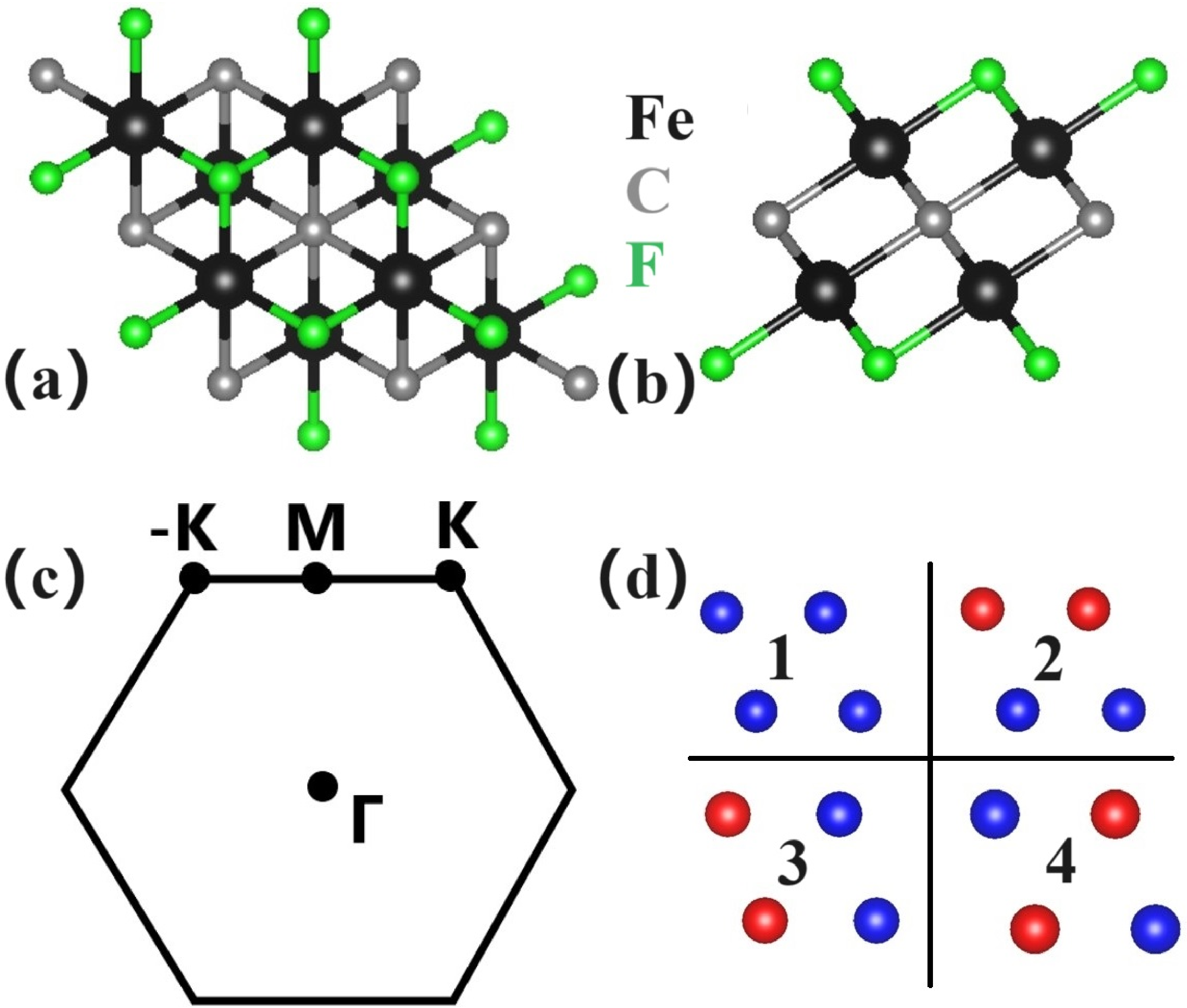}
  \caption{(Color online) For monolayer  $\mathrm{Fe_2CF_2}$,  (a) and (b): the top and side views of crystal structures; (c):  the first BZ with high symmetry points;  (d): four magnetic configurations, including FM (1), AFM1 (2), AFM2 (3), and AFM3 (4) ordering. The blue (red) balls represent the spin-up (spin-down)
Fe atoms.}\label{st}
\end{figure}

For A-type AFM hexagonal monolayer $\mathrm{Cr_2CH_2}$, the biaxial strain need be applied to change its valence band maximum  (VBM)  from $\Gamma$ to -K or K point, and the valley splitting  is about 49 meV\cite{gsd1}. Only out-of-plane magnetic orientation can produce spontaneous valley polarization in  AFM hexagonal monolayer, and the calculated magnetic anisotropy energy (MAE) is 27$\mathrm{\mu eV}$/unit cell\cite{gsd1}.
It is very important and meaningful to search for   AFM hexagonal monolayer with  intrinsic -K/K valley, large valley splitting and strong out-of-plane magnetic anisotropy.
Recently,   2D A-type AFM   $\mathrm{Fe_2CX_2}$ (X = F, Cl) and Janus  $\mathrm{Fe_2CFCl}$
monolayers have been predicted with strong out-of-plane magnetic anisotropy and high N$\acute{e}$el temperatures\cite{zj}, which share the similar crystal structures with $\mathrm{Cr_2CH_2}$.
In this work, we predict that $\mathrm{Fe_2CF_2}$ and $\mathrm{Fe_2CFCl}$ possess large  spontaneous valley polarization, which can be used to achieve AVHE.

\section{Computational detail}
 The spin-polarized  first-principles calculations are performed within density functional theory (DFT)\cite{1} by using the projector augmented-wave (PAW) method,  as implemented in Vienna ab initio Simulation Package (VASP)\cite{pv1,pv2,pv3}.  We use  generalized gradient
approximation  of Perdew-Burke-Ernzerhof (PBE-GGA)\cite{pbe}as the exchange-correlation functional with the kinetic energy cutoff  of 500 eV,  the total energy  convergence criterion of  $10^{-8}$ eV and the force convergence criterion of 0.0001 $\mathrm{eV.{\AA}^{-1}}$. To account for the localized nature of Fe-3$d$ orbitals, a Hubbard correction $U$ is used by  the
rotationally invariant approach proposed by Dudarev et al\cite{du}. In ref.\cite{zj}, the linear response method has been
used to determine the effective $U$ ($U_{eff}$) values, and they are 4.26, 4.96, and 4.27 eV for $\mathrm{Fe_2CF_2}$, $\mathrm{Fe_2CCl_2}$ and Janus $\mathrm{Fe_2CFCl}$ monolayers, which are also adopted in our works.
The SOC is incorporated for investigation of valley splitting and MAE.
A slab model with a vacuum thickness of more than 30 $\mathrm{{\AA}}$ along $z$ direction is used to avoid interlayer
interactions. A 21$\times$21$\times$1 Monkhorst-Pack k-point meshes are used to sample the Brillouin zone (BZ) for calculating electronic structures. The Berry curvatures
are obtained directly from the calculated
wave functions  based on Fukui's
method\cite{bm},  as implemented in  the VASPBERRY code\cite{bm1,bm2}.

\begin{figure}
  % Requires \usepackage{graphicx}
  \includegraphics[width=8cm]{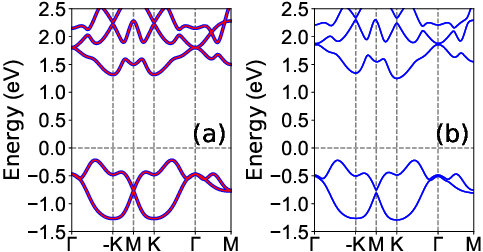}
\caption{(Color online)For  $\mathrm{Fe_2CF_2}$, the energy band structures  without SOC (a) and  with SOC (b) for magnetization direction along the positive $z$ direction.  In (a), the blue (red) lines represent the band structure in the spin-up (spin-down) direction.}\label{band}
\end{figure}

\section{Crystal structures and magnetic ground state}
The crystal structures of $\mathrm{Fe_2CF_2}$ along with the first BZ are  shown in \autoref{st} (a), (b) and (c), which consists of five atomic layers in the sequence of F-Fe-C-Fe-F, namely with the middle C layer
sandwiched between two Fe-F bilayers. The $\mathrm{Fe_2CF_2}$ crystallizes in the  $P\bar{3}m1$ space group (No.~164),  hosting spatial  inversion symmetry without considering magnetic ordering. The magnetic ground state of  $\mathrm{Fe_2CF_2}$ can be determined by comparing the energies of FM and three  AFM (AFM1, AFM2 and AFM3) configurations, as  shown in \autoref{st} (d).  The AFM1 magnetic configuration of them   is called A-type AFM state with the intralayer FM and interlayer AFM couplings. It is proved  that the energy of AFM1 per unit cell is 651 meV, 15 meV and  395 meV  lower  than those of FM, AFM2 and AFM3 cases, confirming that the $\mathrm{Fe_2CF_2}$  possesses A-type AFM ordeing.
When spin is considered for  $\mathrm{Fe_2CF_2}$,  the inversion symmetry $P$ is missing for A-type AFM ordering, and the time-reversal symmetry $T$ is also lacking. However, a combination of inversion symmetry $P$ and time-reversal symmetry $T$  ($PT$) exists in A-type AFM $\mathrm{Fe_2CF_2}$ (see FIG.S1 of electronic supplementary information (ESI)). The optimized  lattice constants are $a$=$b$=3.06 $\mathrm{{\AA}}$ by GGA+$U$ method, which is consistent with the previous  value of 3.03 $\mathrm{{\AA}}$\cite{zj}. The  total magnetic moment of $\mathrm{Fe_2CF_2}$  per unit cell is strictly 0.00 $\mu_B$ with the  magnetic moment of bottom/top Fe atom  being 3.91  $\mu_B$/-3.91 $\mu_B$.
\begin{figure}
  % Requires \usepackage{graphicx}
  \includegraphics[width=8cm]{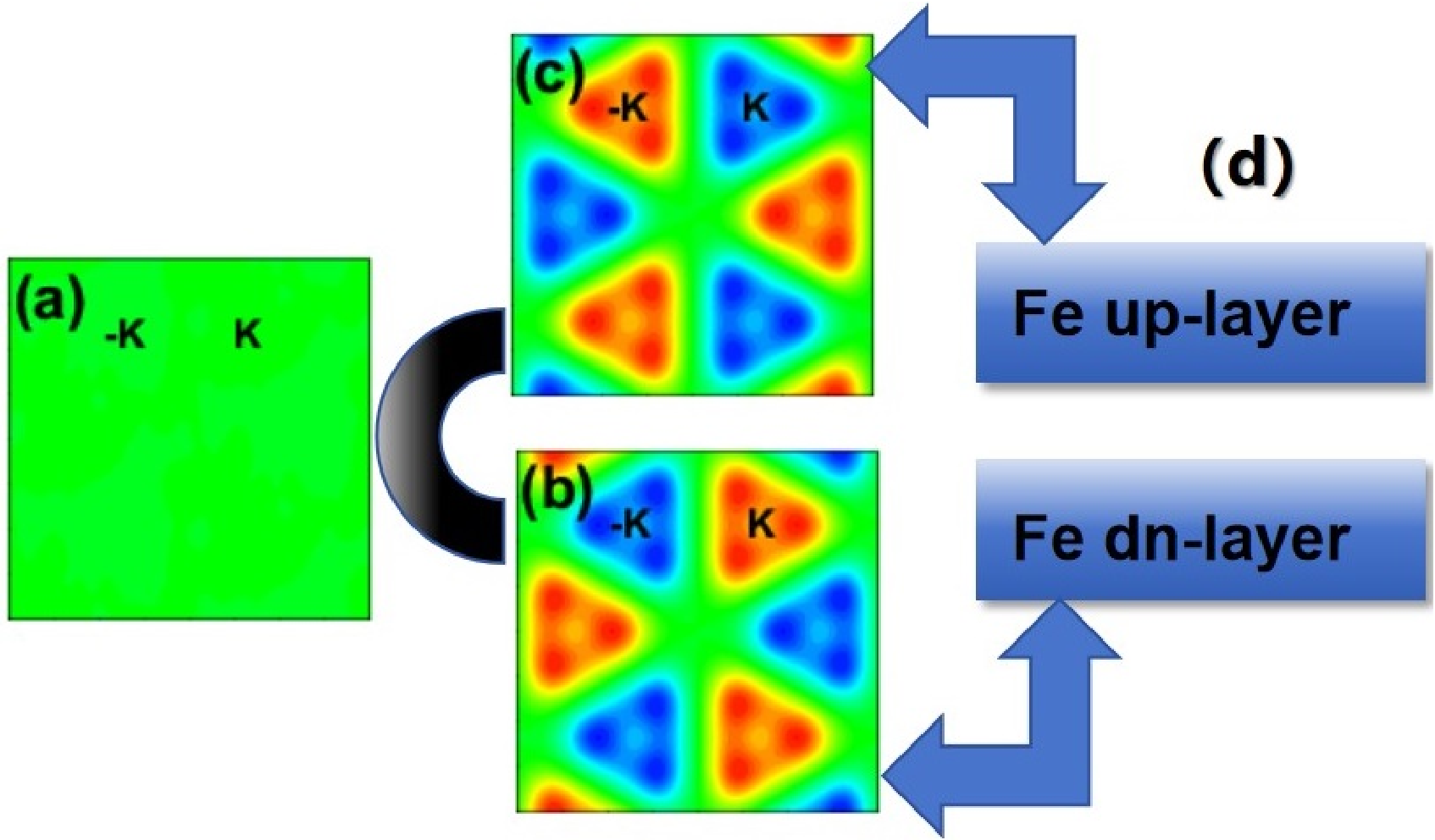}
\caption{(Color online)For  $\mathrm{Fe_2CF_2}$, the distribution of Berry curvatures of  total (a),  spin-up (b) and
spin-down (c). The  $PT$ symmetry leads to net-zero Berry curvature in momentum space, but  the Berry
curvatures for the spin-up and spin-down channels are nonzero with opposite signs, producing  layer-locked hidden
Berry curvature in real space (d).}\label{berry}
\end{figure}

\section{Valley splitting and  hidden Berry curvature}
 The energy band structures of $\mathrm{Fe_2CF_2}$ are plotted in \autoref{band}  without SOC  and  with SOC for magnetization direction along the positive $z$ direction.
Based on \autoref{band} (a), the spin-up and spin-down channels are degenerate due to the $PT$ symmetry, and the $\mathrm{Fe_2CF_2}$  is  an indirect band gap semiconductor with gap value of 1.54 eV. It is clearly seen that the energies of  -K and K valleys in the conduction band are degenerate.
The lattice of  $\mathrm{Fe_2CF_2}$ has inversion symmetry, but  the opposite spin vectors of the two sublattices
break  spatial inversion ($P$) symmetry and time reversal ($T$)
symmetry,  giving rise to spontaneous valley polarization. The broken $P$ and $T$ can also be found in ferrovalley materials with spontaneous valley polarization\cite{q10}.
According to \autoref{band} (b), when the SOC is included, the spin splitting is still absent, but the energy of -K valley
becomes  higher than one of K valley, producing valley polarization with valley splitting of 97 meV ($\Delta E_C=E^{-K}_C-E^{K}_C$).
The valley splitting of  $\mathrm{Fe_2CF_2}$ is higher than those of available monolayer AFM valley materials\cite{v13,v14,gsd1,gsd3}, and even higher than the valley splittings  of many widely studied ferrovalley  materials\cite{q10,q11,q12,q13,q13-1,q14,q14-1,q14-2,q15,q16,q17,q18}.
When including SOC, the $\mathrm{Fe_2CF_2}$  is still an indirect band gap semiconductor, and the gap value is reduced to 1.47 eV.

\begin{figure}
  % Requires \usepackage{graphicx}
  \includegraphics[width=7.0cm]{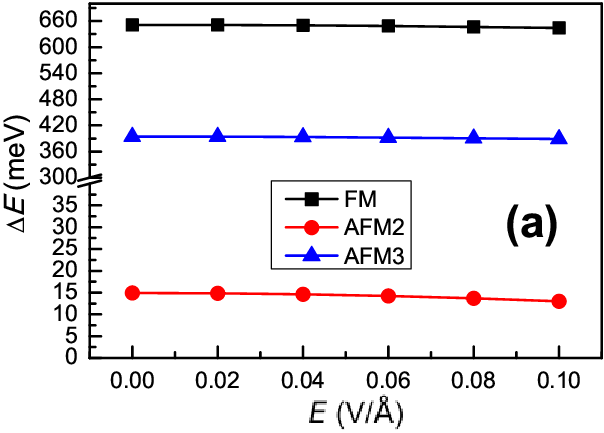}
    \includegraphics[width=7.2cm]{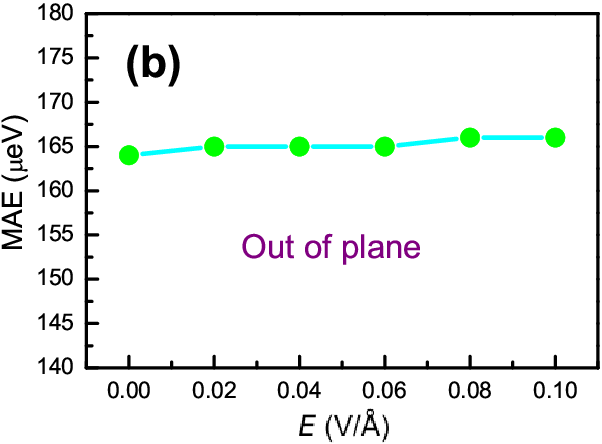}
\caption{(Color online)For  $\mathrm{Fe_2CF_2}$, (a): the energy difference between  FM/AFM2/AFM3 and AFM1 orderings as a function of $E$; (b): the MAE vs $E$.}\label{ene}
\end{figure}

The symmetry of a system can be affected by different magnetic orientation, which has important effects on valley polarization\cite{gsd4,gsd5}.
For example, when the magnetic orientation is out-of-plane, the valley splitting can be observed for magnetic valley materials; while for in-plane magnetization, the spontaneous valley polarization will disappear.
To achieve spontaneous valley polarization, an out-of-plane magnetic orientation is needed, and the MAE can be used to  determine the magnetic orientation, which is defined as $E_{MAE}=E^{||}_{SOC}-E^{\perp}_{SOC}$ where $||$ and $\perp$ denote
the in-plane and out-of-plane spin orientations.
 By GGA+$U$+SOC method, the calculated MAE is 164 $\mathrm{\mu eV}$/unit cell, and the positive value indicates   the out-of-plane easy magnetization axis of $\mathrm{Fe_2CF_2}$.  The out-of-plane easy magnetization axis confirms the  spontaneous valley polarization of $\mathrm{Fe_2CF_2}$.

\begin{figure}
  % Requires \usepackage{graphicx}
  \includegraphics[width=8cm]{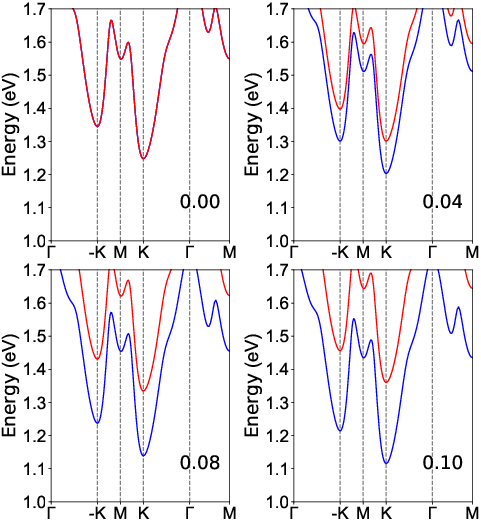}
\caption{(Color online)For  $\mathrm{Fe_2CF_2}$, the spin-resolved conduction bands near the Fermi level with SOC at representative $E$ (0.00, 0.04, 0.08 and 0.10  $\mathrm{V/{\AA}}$).}\label{band-1}
\end{figure}
Due to $PT$ symmetry of $\mathrm{Fe_2CF_2}$, there is zero berry curvature ($\Omega(k)$) everywhere in the momentum space.
However, because each layer unit (F-Fe-C) breaks the local $PT$ symmetry, there is layer-locked hidden Berry curvature\cite{lqh}.   The A-type AFM ordering produces  layer-spin locking,  which makes the Berry curvatures of  spin-up and spin-down channels  equal in magnitude and opposite in sign.
For $\mathrm{Fe_2CF_2}$, the distribution of Berry curvatures of  total,  spin-up  and
spin-down  are  shown in \autoref{berry}.  It is clearly seen  that  the total berry curvature  everywhere in the momentum space is zero, and
the hot spots of spin-resolved Berry curvatures
locate at the -K and K valleys. The berry curvatures  show opposite signs for different valley of  the same spin channel and   the same valley of different spin channel, producing  layer-locked hidden
Berry curvature in real space. By applying a longitudinal in-plane electric field,
the Bloch carriers will acquire an anomalous transverse
velocity $v_{\bot}$$\sim$$E_{\parallel}\times\Omega(k)$\cite{q4}. When
the Fermi level is  shifted  between the -K and K valleys in the conduction band, the spin-up and spin-down carriers from K valley will
accumulate along opposite sides of different layer, resulting
in the valley layer-spin Hall effect (FIG.S2 of ESI), but the AVHE is absent due to spin degeneracy.

\section{Spin splitting caused by electric field and AVHE}
An out-of-plane electric field can be used to break the $PT$ symmetry by removing $P$ symmetry of crystal structure\cite{x1}, and then the removal of
spin degeneracy of -K and K valleys can be achieved.  The layer-dependent electrostatic potential can be induced by an out-of-plane electric field, which leads to spin splitting effect. Here, an out-of-plane  electric field $+E$ (0.00-0.10 $\mathrm{V/{\AA}}$) is applied, and the $-E$  produces  exactly the same results except spin order, which is because two Fe layers with opposite magnetic moments are related by a glide mirror $G_z$ symmetry.  Within applied out-of-plane electric field,  the magnetic ground state  of $\mathrm{Fe_2CF_2}$ is determined by
calculating energy difference between  FM/AFM2/AFM3 and AFM1 orderings. According to \autoref{ene} (a), A-type AFM ordering (AFM1 ordering) is always ground state within considered $E$ range, and the electric field has small influence on magnetic energy differences.
The MAE vs $E$ is shown in \autoref{ene} (b), and  the positive MAE confirms the out-of-plane magnetic orientation of $\mathrm{Fe_2CF_2}$ within considered $E$ range. These  ensure that the AVHE can be achieved in $\mathrm{Fe_2CF_2}$ by electric field engineering.
\begin{figure}
  % Requires \usepackage{graphicx}
  \includegraphics[width=8cm]{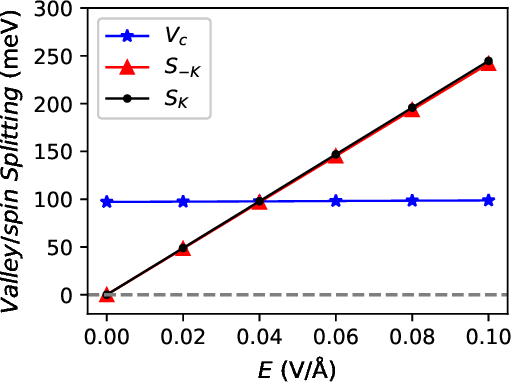}
\caption{(Color online)For $\mathrm{Fe_2CF_2}$, the valley splitting ($V_c$) and spin splitting ($S_{-K}$ and $S_{K}$ at -K and K valleys) of  conduction  band   as a function of $E$. }\label{ss}
\end{figure}

The energy band structures of $\mathrm{Fe_2CF_2}$ at representative $E$ without SOC and with SOC are plotted in FIG.S3 of ESI, and the  enlarged figures of spin-resolved SOC energy band structures near the Fermi level for the conduction band are plotted in \autoref{band-1}.
With applied electric field, it is clearly seen that the spin degeneracy is removed. The valley splitting  and spin splittings of  -K and K valleys in the conduction band  as a function of $E$ are shown in \autoref{ss}.
With increasing $E$, the valley splitting increases from 97 meV to 99 meV, maintaining large spontaneous valley polarization.
It is clearly seen that the spin splittings of -K and K valleys almost coincide, and  show a linear relationship with $E$.
 The out-of-plane electric field induces layer-dependent electrostatic potential $\varpropto$ $eEd$\cite{rz}, where $e$ and $d$ denote the electron charge and the distance of two Fe layers. The spin splittings of -K and K valleys can be estimated by  $eEd$. Taking $E$=0.10$\mathrm{V/{\AA}}$ as a example, the estimated spin splitting   is approximately 233 meV with the $d$ being 2.33 A, which is close to DFT result of 242 meV.
  With the electric field from $+E$ to $-E$, the layer-dependent electrostatic potential is  reversed, which makes spin order  of spin splitting  reverse, but the sizes of spin splitting and  valley splitting  remain unchanged (FIG.S4 of ESI).

\begin{figure}
  % Requires \usepackage{graphicx}
  \includegraphics[width=8cm]{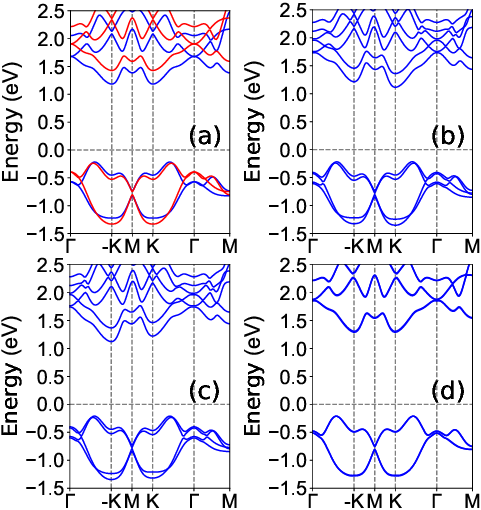}
\caption{(Color online)For  $\mathrm{Fe_2CF_2}$ at $E$=0.10 $\mathrm{V/{\AA}}$, the energy band structures  without SOC (a), and  with SOC (b, c, d) for magnetization direction along the positive $z$, negative $z$, and positive $x$ direction, respectively.  In (a), the blue (red) lines represent the band structure in the spin-up (spin-down) direction.}\label{band-2}
\end{figure}

For  $\mathrm{Fe_2CF_2}$ at $E$=0.10 $\mathrm{V/{\AA}}$, the energy band structures  without SOC and  with SOC  for magnetization direction along the positive $z$, negative $z$, and positive $x$ direction  are plotted in \autoref{band-2}.
According to \autoref{band-2} (a), the
spin splitting can be observed due to the broken $PT$ symmetry, and the energies of  -K and K valleys in the conduction band are degenerate.  \autoref{band-2} (b) shows that the SOC can produce spontaneous valley polarization with valley splitting of 99 meV, and  the energy of -K valley
is higher than one of K valley.  When  the magnetization direction is reversed, the valley polarization can  be
switched with the same size of valley splitting(\autoref{band-2} (c)), and  the energy of K valley
is higher than one of -K valley. When the  magnetization direction of  $\mathrm{Fe_2CF_2}$ is in-plane along $x$ direction (\autoref{band-2} (d)), no valley polarization  and  no obvious spin splitting can be observed. These similar magnetization-direction induced phenomena can also be observed in ferrovalley monolayers\cite{q11,q12,q13,q13-1,q14,q14-1,q14-2,q15,q16,q17,q18}.

\begin{figure}
  % Requires \usepackage{graphicx}
  \includegraphics[width=8cm]{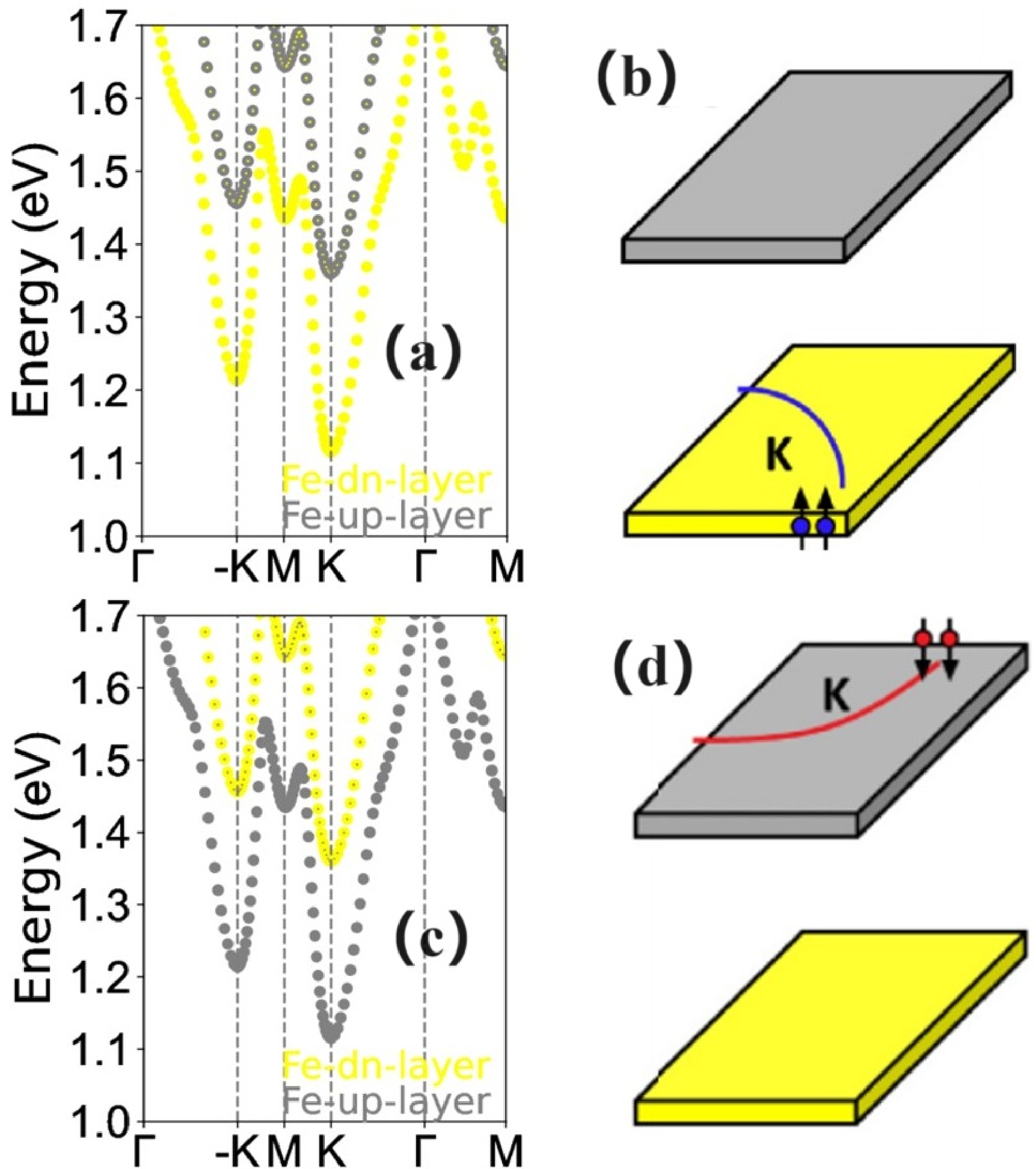}
  \caption{(Color online) For monolayer  $\mathrm{Fe_2CF_2}$,   the Fe-layer-resolved conduction bands near the Fermi level with SOC at $E$=+0.10 (a) and -0.10 (c) $\mathrm{V/{\AA}}$. In the presence of a longitudinal in-plane electric field,  an appropriate electron doping for (a) and (c) produces layer-locked anomalous valley Hall effect (b) and (d). The upper and lower planes represent  the top and bottom Fe layers.}\label{avh}
\end{figure}

For $\mathrm{Fe_2CF_2}$ at $E$=0.10 $\mathrm{V/{\AA}}$, the distribution of Berry curvatures of  total,  spin-up  and
spin-down  are  plotted in FIG.S5 of ESI.  Due to broken $PT$ symmetry, the total nonzero berry curvature  in the momentum space can be observed.
 It is clearly seen that the  Berry curvatures show  opposite sign  for the same valley at different spin channel and different valley at the same spin channel. For monolayer  $\mathrm{Fe_2CF_2}$,   the Fe-layer-resolved conduction bands near the Fermi level with SOC at $E$=+0.10  and -0.10  $\mathrm{V/{\AA}}$ are plotted in \autoref{avh} (a) and \autoref{avh} (c).
  Under an in-plane electric field,  when  the Fermi level is shifted  between the -K and K valleys in the conduction band, only the spin-up carriers from the K valley move to the boundary of  bottom layer of the sample (\autoref{avh} (b)), producing layer-locked AVHE.
  Conversely, the spin-down carriers from the K valley will move to the opposite side of top layer of  the sample (\autoref{avh} (d)), when the direction of  electric field is reversed.  This transverse accumulation of spin-polarized carriers can give rise to  a net charge/spin current, and \autoref{avh} (b) and \autoref{avh} (d) generate opposite measurable voltage.

\section{Spin splitting caused by Janus structure}
\begin{figure}
  % Requires \usepackage{graphicx}
  \includegraphics[width=8cm]{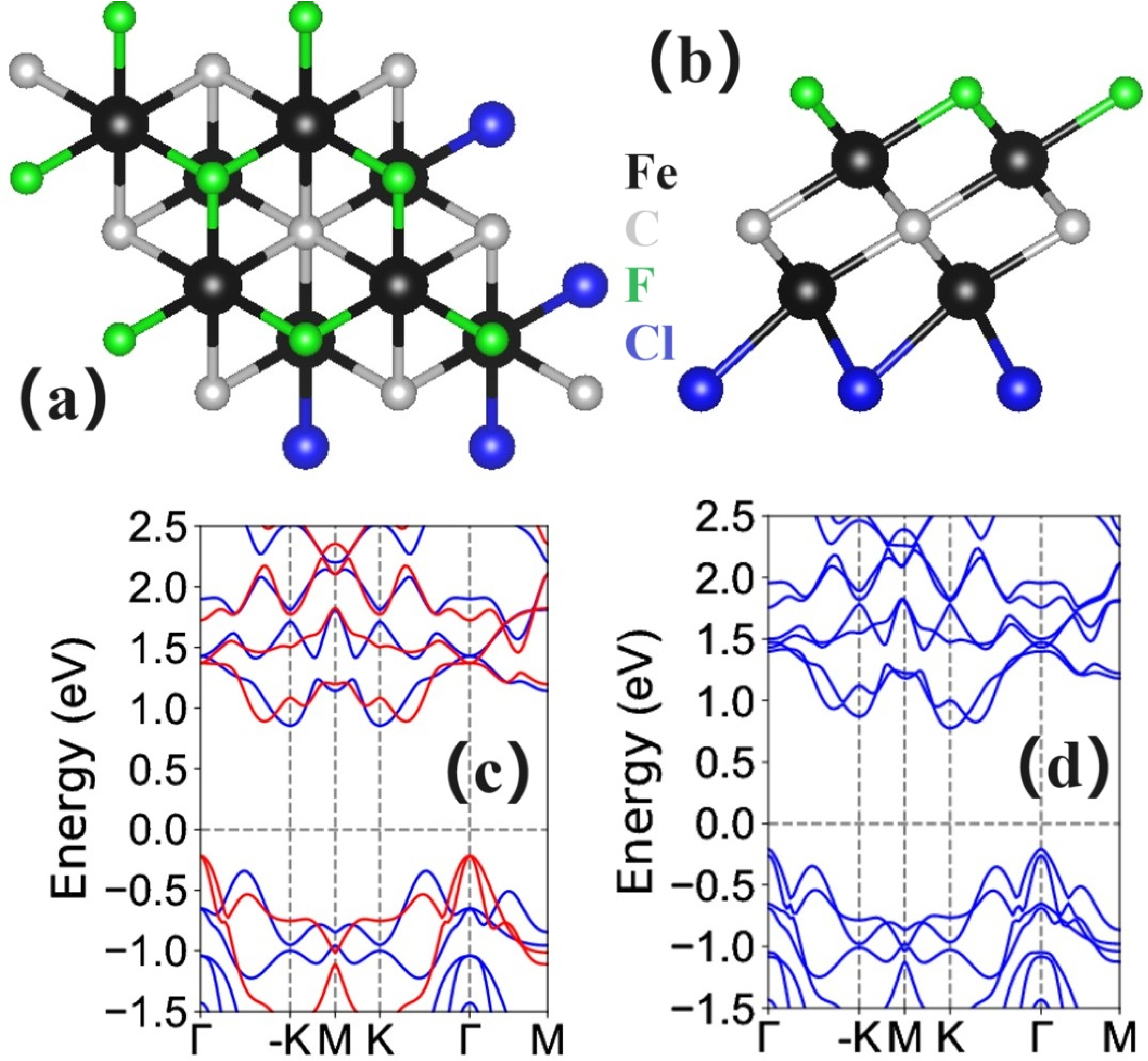}
  \caption{(Color online)  For Janus monolayer  $\mathrm{Fe_2CFCl}$,  (a) and (b): the top and side views of crystal structures; the energy band structures  without SOC (c) and  with SOC (d) for magnetization direction along the positive $z$ direction.  In (c), the blue (red) lines represent the band structure in the spin-up (spin-down) direction.}\label{st-1}
\end{figure}

The spontaneous valley polarization and electric-field induced spin splitting can also be achieved in  $\mathrm{Fe_2CCl_2}$ (see FIG.S6 of ESI), but these valleys in the conduction band  deviate from high symmetry -K/K point.
For Janus monolayer, an out-of-plane intrinsic polar electric field can appear, which is equivalent to an external electric field for producing spin splitting\cite{gsd3}.
By replacing the top F layer by Cl element in $\mathrm{Fe_2CF_2}$, the Janus monolayer $\mathrm{Fe_2CFCl}$  is obtained (see \autoref{st-1} (a) and (b)), which has the symmetry of $P3m1$ (No.~156) without considering spin, lacking inversion symmetry $P$.
Calculated results show that the energy of AFM1 per unit cell is 648 meV, 15 meV and  373 meV  lower  than those of FM, AFM2 and AFM3 cases, confirming that the $\mathrm{Fe_2CFCl}$  has A-type AFM ordering. When considering spin,  the inversion symmetry $P$ and time-reversal symmetry $T$ are missing for A-type AFM $\mathrm{Fe_2CFCl}$, and  a combination of inversion symmetry $P$ and time-reversal symmetry $T$  ($PT$) also lacks. According to the the plane-averaged electrostatic potential along the $z$ direction, the built-in electric field  of  $\mathrm{Fe_2CFCl}$ is predicted to be 1.71 $\mathrm{V/{\AA}}$ (see FIG.S7 of ESI). In fact, Janus monolayer $\mathrm{Fe_2CFCl}$ is an electric-potential-difference AFM material\cite{gsd3}, which possesses spontaneous spin splitting.

To confirm the  spontaneous valley polarization of $\mathrm{Fe_2CFCl}$, the MAE is calculated, and the predicted value  is 166$\mathrm{\mu eV}$/unit cell, which means that the  easy magnetization axis of $\mathrm{Fe_2CFCl}$ is out-of-plane.
 The energy band structures of $\mathrm{Fe_2CFCl}$ are plotted in \autoref{st-1} (c) and (d)  without SOC  and  with SOC for magnetization direction along the positive $z$ direction.  Due to the broken $PT$ symmetry,
 the obvious spin splitting can be observed in $\mathrm{Fe_2CFCl}$, and it is  an indirect band
gap semiconductor without valley polarization (\autoref{st-1} (c)).  When considering the SOC,
\autoref{st-1} (d) shows spontaneous valley polarization with a valley splitting of  95 meV, which is larger than ones of many magnetic valley materials\cite{v13,v14,gsd1,gsd3,q10,q11,q12,q13,q13-1,q14,q14-1,q14-2,q15,q16,q17,q18}.
The  total magnetic moment of $\mathrm{Fe_2CFCl}$  per unit cell is strictly 0.00 $\mu_B$, and the  magnetic moment of bottom/top Fe atom  is 3.82  $\mu_B$/-3.90 $\mu_B$.

For $\mathrm{Fe_2CFCl}$, the distribution of Berry curvatures of  total,  spin-up  and
spin-down  are  shown in FIG.S8 of ESI.  Due to broken $PT$ symmetry, $\mathrm{Fe_2CFCl}$ shows the total nonzero berry curvature  in the momentum space, and opposite sign  for the same valley at different spin channel and different valley at the same spin channel.
   By shifting the Fermi level between the -K and K valleys in the conduction band, the spin-up  carriers from K valley will
accumulate along the bottom boundary of the sample under a longitudinal in-plane electric field, resulting
in the layer-locked AVHE. Therefore, for Janus $\mathrm{Fe_2CFCl}$, the AVHE can be achieved without external electric field.

\section{Conclusion}
In summary,  we present an A-type hexagonal AFM monolayer $\mathrm{Fe_2CF_2}$ to realize AVHE  by applying out-of-plane electric filed.
 The spontaneous valley polarization can  occur in $\mathrm{Fe_2CF_2}$ with the valley splitting of about 97 meV due to intrinsic the out-of-plane magnetic orientation, but the spin splittings of -K and K valleys are absent, preventing AVHE.  The introduction of
an out-of-plane electric field induces the  spin splitting  in monolayer $\mathrm{Fe_2CF_2}$, which is  due to layer-dependent electrostatic potential.  By combining with layer-locked hidden Berry curvature, the layer-locked AVHE can be achieved in monolayer $\mathrm{Fe_2CF_2}$ by applying external electric field.
Our works enrich valleytronic materials with  AFM ordering, which  provides  advantageous for the development of energy-efficient and ultrafast electronic devices.

\begin{acknowledgments}
This work is supported by Natural Science Basis Research Plan in Shaanxi Province of China  (2021JM-456). We are grateful to Shanxi Supercomputing Center of China, and the calculations were performed on TianHe-2..
\end{acknowledgments}

\end{document}